\documentstyle[emlines2,tighten,eqsecnum,aps]{revtex}

\newtheorem{theorem}{Theorem}
\newtheorem{suggestion}[theorem]{Suggestion}

\begin{document}
\title{Mechanism of Self-regulation \\
in a Simple Model of Hierarchically Organized Market }
\author{V.~V.~Gafiychuk}
\address{Institute for Applied Problems of Mathematics and Mechanics, National\\
Academy of Science of Ukraine 3b, Naukova St. L'viv, 290601, Ukraine.}
\author{I.~A.~Lubashevsky, Yu.~L.~Klimontovich}
\address{Moscow State University, Department of Physics, Vorobievy Gory, Moscow\\
119899, Russia. }
\date{\today}
\maketitle

\begin{abstract}
We propose a model for a market which structure is of the tree form. Each
branch of the tree is composed by identical firms, its root (the branch of the
first level) is formed by the firms producing raw material, and the branches
of the last level are the retail outlets. The branching points (tree nodes)
are micromarkets for firms forming branches connected with a given node. The
prices and the production rate are controlled by the ballance in supply and
demand, the competition is assumed to be perfect. We show that such a market
functions perfectly: the prices are specified by the production expense
only, whereas demand determines the production rate.

We construct an efficiency functional which extremal gives the governing
equations for the market. It turns out that this ideal market is degenerated
with respect to its structure. It is shown, that such market functions
ideally: the prices are determined by the costs on production of the goods,
and the level of production of the goods of any kind defined only by demand
for the goods of this sort.
\end{abstract}

\section{\protect\bigskip Introduction\label{sec:intr}}

\bigskip The characteristic properties of the commodity market evolution
have been intensively studied during recent years. However in the present time
theoretical description of self-organization phenomena in microeconomic
market is still a practicaly unresolved challenging problem. The reason is that the
commodity market is a radically open system, i.e. such a system, which exists
while and if it exchanges by energy, substance and, that is of not less importance,
by information with the surrounding world (with the social and
physical environment). This is the main difference between market
systems and physical systems. Really, for physical systems it is possible to
indicate a state of a thermodynamic equilibrium, where all streams are equal
to zero. Such equilibrium state is usually described with small amount of
macroscopic parameters - parameters of ``order''. But economic systems exist
while they have the streams of commodity and money. Besides, the number of
parameters, which define their state at first glance, seems to be actually
endless.

Nevertheless, we can consider far from equilibrium dissipative systems
found in physics and chemistry as some rough analogs of such economic systems
from the viewpoint of common features of evolution. During recent twenty years
these dissipative systems were profoundly investigated (see, for example, 
\cite{NP79r,H80r,H91r,K95r,ker}). It has been found, that their processes
of dissipation can produce the self-organization in various types of 
dissipative structures.

It should be noted, that physical or chemical system at a moment of the
origin of dissipative structures transforms into a state with an intermediate
degree of chaosity \cite{K95r}. The entropy of this state is less maximum,
corresponding to thermodynamic equilibrium. At the same time it is possible
to describe the state with a set of macroscopic (or mesoscopic) parameters
of order. Due to the slow change of these parameters the evolution of such
systems is usually realized as the process of transition through the
sequence of nonequilibrium quasistationary states, distinguished from each
other by their dissipative structures.

In economic systems also appear the complex dissipative structures
stipulated by unaware will of people. This structures arise from individual
interaction of a plenty of agents and spontaneous formation of some order in
their relationships (for the review of this problem from the viewpoint of a
synergetics, see, for example, \cite{H92r,W91}).

As an example of such self-organization phenomena one can consider
the creation and function of trade networks in the systems with distributed
auctions, investigated in papers \cite{B77,RW85,W90,RG94,GSh95r}.
It had been supposed, that the number of trade agents is beforehand given
and would not vary in time. It should be noted that self-organization
and evolution of these structures is similar to nonequilibrium
phase transitions under the changes of external parameters \cite{GSh95r}.

In the first stage of the development the self-organization theory in
microeconomics it is necessary to choose some adequate zero approximation,
which actually is a detailed description of the market ideal state - its
``norm''. Such norm is well-known. It is an equilibrium market, where the
interests of buyers and sellers completely coincide with each other in that
way, that within the given price the value of supply is equal to the value of
demand. However, for the analysis of self-organization processes more
detailed description is required, for example, to pair interactions of
sellers and buyers.

Apparently, such rather detailed description of economics as a whole does not have
much sense, because the markets of radically different products would be
slightly associated. For example, the market of meat items and the market of
furniture can interact between themselves mainly through changes in financial
states of the whole collection of the consumers. It should be noted, that
such situation is similar to the situation, realized in living organisms.
Indeed, if a living organism is not in extreme conditions, each
of its organs is supplied with amount of blood, completely satisfying its needs,
regardless of the other organ's function. It is supplied by the system of
large arteries. These arteries form the infinite tank of blood for system of
separate organs, and the heart replies to some aggregated information about
the state of organism as a whole \cite{M89r}.

Therefore it would be worthwhile to restrict our consideration to any
commodity market, related from the viewpoint of possibility of their mutual
substitution. Producers, dealers and consumers of these goods will form a
common connected network, carrying on the goods from the producers of raw
material up to the buyers which buy these goods in the retail shops.

Nevertheless, the set of products, made on the given market, form
only a small part of the whole public market. {\it \ }The state of the rest
of the economic system has to be taken into account in some set of
governing parameters, which actually do not influence the local changes on
the given market and reflect aggregated information about society as a whole.

In other words, for the analysis of self-organization processes it seems
reasonable to restrict ourselves to research of some small markets of goods,
which nevertheless form a common connected system between consumers and all
producers of these goods. Such mesoscopic market (mesomarket) is a subject
of research in the present paper. Concerning the common requirements applied
to the possible model of its norm, it is worhty to single out the following
items.

{\it First}, it should be of an intermediate degree of chaosity. In other
words , on the one hand, individual behavior of the participants of ideal market
is not prescribed from any center. More likely, it is determined by their own
goals, based on the small part of information about the market state. On
the other hand, the ideal market contains structures, generated by individual
interaction of the participants as a result of the establishment of spontaneous order
in their ratios. These structures form the streams of goods and corresponding
flow of money in the opposite direction.

These streams of money bear some kind of information self-processing,
which provides a way for each participant to react adequately
on various changes of market state, basing only on a small share of
information, which aggregates the information about market state as a whole
(see \cite{LG95ES,gaf1,gaf2}).

It should be noted, that strictly administrative (completely ordered)
system, based on the ideal overall plan as well as formally equivalent to it
Walras' ''auctioneer'' (completely unregulated system, where everyone can
contact  each other through some process such as\ ''tatonment'') cannot
function in reality, because for proper functioning these systems require physically
infinite time to collect the information\cite{P88r,P89r}.

Summing up this item we can formulate, that a model of norm has to contain
individual interaction of the participants, described with classical laws of
demand - supply. And the realization of their contacts has to be given by
the micromarket network structure, where physically finite amount of the
participants comes into contact.

{\it Second} the prices, arising in the result of individual trade relations,
normally have to be determined only by real material costs on commodity
production. It should be noted, however, that this correspondence should be
not an outcome of the prices control, but a result of self-organization processes.

{\it Third}, spontaneously arising structures should be organized
hierarchically. This is due to the fact of the existence of huge amount of
various goods in the market in contrast to small amount of raw materials.
Therefore, before the final goods will be originated , material, of which
they are made, will be processed during the stages of their production for
many times. Their sequence and content are determined by the sort of the
produced product and are various for different types of products. The simple
example of such hierarchical structure is the structure of the tree form.
The root of this tree is formed by firms, obtaining raw materials, and the branches of
the last level are the retail shops.

{\it Fourth}, the ideal market should be characterized by separability
concerning the production of the commodity. In other words, let $\{\alpha \}$
is a collection of the final types of goods of the given market, $%
\mbox{\boldmath$p$}=\{p_{\alpha }\}$ is their price in retail shops
and $\mbox{\boldmath$S$}(\mbox{\boldmath$p$},\phi )=\{S_{\alpha }(%
\mbox{\boldmath$p$},\phi )\}$ - their demand function, i.e. a consumption
level at a given prices. The demand function depends also on some internal
parameters $\phi $, describing the state of consumers, for example, their
average yearly income and so forth. Then the change in the
demand for the goods of sort $\alpha $ (i.e. change of the function $%
S_{\alpha }(\mbox{\boldmath$p$},\phi )$, for example, as a result of
variations in $\phi $) parameters should not affect prices and the
level of production of the other types of goods $\alpha ^{\prime }\neq
\alpha $ under their invariable demand (i.e. in the absence of changes in
the function $S_{\alpha ^{\prime }}(\mbox{\boldmath$p$},\phi )$).

Really, in an ideal case the consumption level of the goods of sort $\alpha $
and their price $p_{\alpha }$ are the result of the balance of total costs
and utility of their production. Therefore, if the given balance is
disturbed only because of a change in demand for the goods of the other sort $%
\alpha ^{\prime }\neq \alpha $ without technology change, then it points to
an imperfection of the market and can result in chaos in the commodity
production. The same is true for one sort of goods, if customers derive
various groups, separated from each other in space so, that they are
supplied with various market elements.

It should be noted, that the given requirement of separability of the ideal
market is a typical condition, imposed on the process of commodity
production and on properties of separate producers one with another interaction.
This requirement does not impose any restrictions on possible
immediate interrelations in demand of the consumers for the goods of various
sort. Such interrelations can arise, for example, because of one individual
is a consumer of different types of goods .

In the present paper we have formulated a simple model of the commodity
market, which functions ideally. In this case we consider, that the
structure of the market has a form of a tree, and competition is perfect.

\section{Model\label{sec:model}}

Let us now consider the market of some goods, made of one sort of raw
material. Collection of the consumers ${\cal M}$ of these goods we shall
present as a union 
\begin{equation}
{\cal M}=\bigcup_{\alpha }{\mbox{\boldmath$m$}}_{\alpha }  \label{M}
\end{equation}
of various groups ${\mbox{\boldmath$m$}}_{\alpha }$. The difference of these
groups is defined in such a way that they consume the different sort of
goods, or that they acquire formally the common sort of goods, but are
supplied by different branches of the market structure.

It should be noted, that in the first case groups ${\mbox{\boldmath$m$}%
_{\alpha }}$ and ${\mbox{\boldmath$m$}_{\alpha ^{\prime }}}$ (for $\alpha
\neq \alpha ^{\prime }$) can be physically derived by the same collection of
people. In the second case - different consumers are separated either on
space, or belonging to various strata of society.

Let us suppose, that the structure of the given market ${\cal N}$, formed
with industrial and commercial network, has a form of a tree (Fig. ~\ref{F.1}%
). The branch $i$ of this tree is a collection of $n_{i}$ independent firms,
bringing out the products of sort $i$, which buy the products from firms,
belonging to the lower level of the network hierarchy ${\cal N}$, and sell
the results of their activity to the firms of higher level. Firms, forming
the common branch $i$, are considered as identical from the viewpoint of
technological process and their trade relations. The root of the tree ${\cal N}$
(the branch of the first level) is formed by the firms, obtaining and
processing raw material. The tree branches of the last level are the
''points'' of retail trade, supplying only one group of consumers. Besides
we suppose, that for any branch $i$ the number of firms $n_{i}$, belonging
to this branch is a large value 
\begin{equation}
n_{i}\gg 1\qquad \forall i.  \label{nl}
\end{equation}

The equation ~(\ref{nl}) allows us to consider the values $\{n_{i}\}$ as
continuous variable $\ (n_{i}+1\approx n_{i})$.

Each network node ${\cal N}$, for example, node ${\cal B}$ is a micromarket,
in which only those firms, which belong to the branch \{$i_{in}^{{\cal B}}\}$%
, participate going in a node ${\cal B}$, and firms, belonging to branches \{%
$i_{out}^{{\cal B}}$\}, going out of the given node (sellers and buyers
in this market, respectively, Fig. ~\ref{F.a}). All products on the given
branch are sold under one price $p_{{}_{{\cal B}}}$. Firms, which belong to
the root of a tree ${\cal N}$, extracting raw material, and the firms of the
last level of a hierarchy sell the goods directly to the consumers.

The network ${\cal N}$ actually determines the economic ratios between the
market participants and sets interrelations of product flows. We shall
consider the situations, when the market of the considered goods exists, i.
e. all the streams should be more than zero.

Let us measure the level of the firm production, which belongs to the branch 
$i$ in a unites of the initial raw material flow $x_{i}$, ``flowing
through'' the given firm. It should be noted, that in this system of unites
the value $x$ having dimensionality ({\it material} \/)/({\it time}\/), and
index $i$ really determines the sort of production of the considered firm.
The full products stream of $X_{i}$can be represented 
\begin{equation}
X_{i}=n_{i}x_{i}.  \label{1}
\end{equation}
Due to the material conservation law, in network nodes ${\cal N}$ relations 
\begin{equation}
X_{i_{in}^{{\cal B}}}=\sum_{j\in \{i_{out}^{{\cal B}}\}}X_{j}.  \label{2}
\end{equation}
are fulfilled for any node ${\cal B}$.

The whole production of the last level firms is acquired by the consumers.
In doing so, for a branch $i_{\alpha }$, supplying a group of the consumers $%
{\mbox{\boldmath$m$}_{\alpha }}$ we can write the following: 
\begin{equation}
X_{i_{\alpha }}=X_{\alpha },  \label{3}
\end{equation}
where $X_{\alpha }$~-- consumption level of the goods by group ${%
\mbox{\boldmath$m$}_{\alpha }}$, expressed in raw material flow unites.

It should be noted, that produced goods on this mesomarket, can include not
only the raw material but also other additional materials. The latest,
however, are acquired by firms individually and their costs are included
into the cost of production. The raw material of the considered model is
like a binding of the different producers in the single network, that is
expressed in the conservation laws (\ref{2}).

The trade interaction on the micromarket ${\cal B}$ results in the origin of
interchanged production price. Taking into account the preceding, we shall
measure the interchanged production by the effective price $p_{{}_{{\cal B}}}
$ of the raw material unit. Then in the result of its activity each firm,
which belongs to the branch $i$, obtains the unit time profit $\pi _{i}$,
which is equal to:

\begin{equation}
\pi _{i}=(p_{i}^{(s)}-p_{i}^{(b)})x_{i}-t_{i}(x_{i}).  \label{4}
\end{equation}

Here $p_{i}^{(s)}$ and $p_{i}^{(b)}$ are prices on the micromarkets, where
firms $i$ represent itself as the seller and buyer, respectively, $t_{i}(x)$%
~-- it's total costs for production with $x_{i}$ level. Following the
standard view, we suppose, that $t_{i}(x)$ is increasing convex function,
growing faster than $x^{1+\epsilon }$ (where $\epsilon $ - some positive
constant). Besides, we consider that

\begin{equation}
t_{i}(0)>0\qquad \forall i.  \label{t0}
\end{equation}

It should be noted, that the nonequality ~(\ref{t0}) is a condition, imposed
on the specific production costs, i.e. production costs, referred to one
firm, considered as indivisible. Total costs $T_{i}$, connected with the
activity of all firms, which form the branch $i$, are equal $%
T_{i}=n_{i}t(x_{i})$ and depend on two arguments: $n_{i}$ and $x_{i}$. If
the value $x_{i}$ is considered as the given parameter of production, then
by virtue of ~(\ref{1}) we can express $T_{i}=X_{i}t_{i}(x_{i})/x_{i}\ $ and
therefore $T_{i}\rightarrow 0$ at $X_{i}\rightarrow 0$, since the condition (\ref
{nl}) allows to consider $n_{i}$ as a continuous variable.

For firms, obtaining raw material we have the expression

\begin{equation}
\left. p_{i}^{(b)}\right| _{i\in root}=0.  \label{r}
\end{equation}

We suppose, that the individual goal of each firm is reaching the maximum
profit. In this case the strategy of the firm $i$ we can describe as the
productions with the $x_{i}$ level, satisfying the following condition

\begin{equation}
\frac{\partial \pi _{i}(x_{i})}{\partial x_{i}}=0.  \label{5}
\end{equation}

The change in the demand, in particular, causes the change in the goods production
level, and therefore, results in the value of gained profit. The latter, in its turn,
stimulates or suppresses activity of the firms. In the same case this
stipulates the arising or vanishing of the firms in the market. If a
competition is perfect (i.e. when there are no barriers for the entrance of
new firms into the existing market) at the equilibrium status, the profit
received by the firm, taking into account all costs, should be equal to zero
(see, for example, \cite{BH88ec,SR90}). Assuming this condition executed,
the value $x_{i}$ should satisfy the conditions:

\begin{equation}
\pi _{i}(x_{i})=0.  \label{6}
\end{equation}

The consumers, purchasing the goods in the ''points'' of retail trade, try
to maximize their utility. Taking into account their utility and budget
constrains we shall describe the behavior of the consumers group $%
\mbox{\boldmath$m$}_{\alpha }$ by a positive definite demand function $%
S_{\alpha }(\mbox{\boldmath$p$},\phi )$

\begin{equation}
X_{\alpha }=S_{\alpha }(\mbox{\boldmath$p$},\phi ),  \label{7}
\end{equation}
where $\mbox{\boldmath$p$}=\{p_{\alpha }\}$ - collection of prices of all
final goods in the shops of retail trade, and the parameters $\phi $ reflect
total expenditure. The equations ~ (\ref{1}) - (\ref{3}) and (\ref{r}) - (%
\ref{7}) describe the functioning of the given market and, in particular,
determine the amount of firms $\{n_{i}\}$, participating in production and
trade of commodities.

Let us analyze characteristics of the formulated model.

\section{ Ideal self-regulation\label{sec:perf}}

At the given structure of the micromarkets the unknown variables are: level
of separate firms production$\{x_{i}\}$, their number $\{n_{i}\}$ in
different branches and prices on the micromarkets $\{p_{{\cal B}}\}$.

The expression~(\ref{4}) for the profit $\pi _{i}$ of the firm $i$ contains
only two independent variables, the difference $(p_{i}^{(s)}-p_{i}^{(b)})$
and the level of production $x_{i}$. Therefore, on the accepted assumptions
about type of functions $t_{i}(x)$ the system of two equations~(\ref{5}), (%
\ref{6}) has a single solution $x_{i}^{\ast }$, being the root of the equation

\begin{equation}
\left. \frac{d\ln t_{i}(x)}{d\ln x}\right| _{x=x_{i}^{\ast }}=1.  \label{10}
\end{equation}
To the rate of production $x_{i}^{\ast }$ corresponds the difference of
prices 
\begin{equation}
\Delta p_{i}\stackrel{{\rm def}}{=}(p_{i}^{(s)}-p_{i}^{(b)})=t_{i}^{\prime
}(x_{i}^{\ast }).  \label{11}
\end{equation}
The outcome~(\ref{11}) allows immediately to find the price $p_{{\cal B}}$,
established on the micromarket (node) ${\cal B}$. Really, considering the
condition~(\ref{r}) and summing up expression~(\ref{11}) along the path $%
{\cal P_{B}}$, leading from the root of the tree ${\cal N}$ to the given
node ${\cal B}$, ( Fig. ~\ref{F.1}) we have the following expression for the
price 
\begin{equation}
p_{{}_{{\cal B}}}=\sum_{i\in {\cal P_{B}}}t_{i}^{\prime }(x_{i}^{\ast }).
\label{12}
\end{equation}
Selecting, in particular, micromarket of the last level $\alpha $, i.e. the
node, linking the firms of the last level and the group of the consumers $%
\mbox{\boldmath$m$}_{\alpha }$, we find the prices for goods of sort $\alpha 
$

\begin{equation}
f_{\alpha }\equiv p_{\alpha }=\sum_{i\in {\cal P}_{\alpha }}t_{i}^{\prime
}(x_{i}^{\ast }).  \label{13}
\end{equation}
Thus we arrive to the following conclusion:

\begin{suggestion}
\ In the given model with the fixed structure ${\cal N}$\ the prices $%
\{p_{{}_{{\cal B}}}\}$, established on the micromarkets ${\cal N}$, are
determined only by technological process, and not by demand of the
consumers.\bigskip 
\end{suggestion}

This is a direct consequence of a perfect competition. In other words, at the
perfect competition, the prices reflect only the costs of the goods
production in spite of the fact that physically they are established as the
result of supply and demand balance. It should be noted, that this result is
similar to that, received in Leontjev models.

Full stream of production $X_{i}$, ``flowing'' through the branch $i$,
reasonably depends on demand. Thereby, in the given model just an amount of
firms $\{n_{i}\}$, forming the branches $\{i\}$ of the market structure $%
{\cal N}$ is controlled by demand of the consumers ${\cal M}$. In other
words, the change of values $\{n_{i}\}$ is a manifestation of
self-regulationts of this type of market.

As seen from conditions~(\ref{3}) and (\ref{7}), total stream of the goods $%
X_{\alpha }=X_{i_{\alpha }}$, made by firms of the last level, which form,
for example, the branch $\alpha $, is determined by 
\begin{equation}
X_{\alpha }=S_{\alpha }(\mbox{\boldmath$f$},\phi ),  \label{14}
\end{equation}
where $\mbox{\boldmath$f$}=\{f_{\alpha }\}.$

The conservation laws ~(\ref{2})of the material streams on the micromarkets
for an arbitrary branch $i$ allow us to write an expression: 
\begin{equation}
X_{i}=\sum_{\alpha \in {\cal M}_{i}}X_{\alpha }  \label{15}
\end{equation}
where ${\cal M}_{i}$ is a collection of groups of the consumers, connected
directly with the given branch $i$ through the higher than the branch $i$ hierarchy
level branches (Fig. ~\ref{F.1}). 
\begin{equation}
n_{i}=\frac{1}{x_{i}^{\ast }}\sum_{\alpha \in {\cal M}_{i}}S_{\alpha }(%
\mbox{\boldmath$f$},\phi ).  \label{16}
\end{equation}

In the conclusion to this part of paper, it should be noted, that the
equation~(\ref{14}) testifies about separability of the given model of the
market from the viewpoint of the absence of mutual influence of the
consumers on different types of commodity. In other words\bigskip

\begin{suggestion}
\bigskip\ Any kind of changes in demands of one group of the consumers $\mbox{\boldmath$m$}%
_{\alpha }$ for the goods of sort $\alpha $ in any way does not affect
the price and production of the other sort of goods $\alpha
^{\prime }\neq \alpha .$
\end{suggestion}

\section{\protect\bigskip Variational principle for the ideal market model%
\label{sec:degen}}

The set of values $\{x_{i}\}$, $\{n_{i}\}$ and $\{p_{{\cal B}}\}$ actually
gives all main conditions of the production and trade on the given
commodities market. In other words these parameters are parameters ''of
order''. The strategy of firms and individual behavior of the consumers
(correlation(\ref{5}), (\ref{6}) and (\ref{7})) determines concrete values
of these parameters. Nonetheless, the formulated model of the market
contains an implicit image of one more parameter of the order, namely,
structure of the local micromarkets, which defines the order of firms trade
interactions. Really, with organization of the commodity market of the given
type, initially fixed is the multitude of the consumers ${\cal M}$
and the source of raw material. Realization of technological process and
commercial network is a \ concrete solutions of the problem on
supplying consumers with the required goods. Solution of the given problem,
basically, can be ambiguous and in this case market self-organization
process has to select this realization of the network ${\cal N}$.

Next paragraph is devoted to the analysis of the ideal market model from
this point of view.

\subsection{ Global function of consumption utility}

\bigskip \bigskip At the beginning we shall
construct some function ${\cal U}%
(\mbox{\boldmath$X$})$ from $\mbox{\boldmath$X$}=\{X_{\alpha }\}$, extremum
properties of which define the function of demand of the consumers $%
\mbox{\boldmath$S$}(\mbox{\boldmath$f$},\phi )$ =  $\{ S_{\alpha }(%
\mbox{\boldmath$f$},\phi ) \}$. Let us suppose that collection of the
consumers ${\cal M}$ consists of $\{{\cal M}_{k}\}$ individual or various
groups, considered as separated individuals 
\begin{equation}
{\cal M}=\bigcup_{k}{\cal M}_{k}.  \label{v1}
\end{equation}

A full set of goods, necessary for the life activity of each of them, is $%
\{\gamma \}=\{\alpha \}\bigcup \{\beta \}$, where, as was used earlier, $%
\{\alpha \}$ are goods, delivered on the considered market and $\{\beta \}$ are
goods, made on other markets. In this part of the paper we shall suppose
that appropriate prices ${\mbox{\boldmath$c$}=(\mbox{\boldmath$p$},%
\mbox{\boldmath$q$})}$ for the goods $\{\gamma \}$ are given. Here, following
the previous paragraphs, we used the denotations $\mbox{\boldmath$c$}%
=\{c_{\gamma }\}$, $\mbox{\boldmath$p$}=\{p_{\alpha }\}$ and $%
\mbox{\boldmath$q$}=\{q_{\beta }\}$.

\bigskip Individuals $\{{\cal M}_{k}\}$ are independent. Each individual \ \ $%
k$ on gaining a set of the goods $\mbox{\boldmath$Z$}^{(k)}=(%
\mbox{\boldmath$X$}^{(k)},\mbox{\boldmath$Y$}^{(k)})$ (where $%
\mbox{\boldmath$Z$}^{(k)}=\{Z_{\gamma }^{(k)}\}$, $\mbox{\boldmath$X$}%
^{(k)}=\{X_{\alpha }^{(k)}\}$ and $\mbox{\boldmath$Y$}^{(k)}=\{Y_{\beta
}^{(k)}\}$) maximize his utility function $U_{k}(\mbox{\boldmath$Z$}%
^{(k)})=U_{k}(\mbox{\boldmath$X$}^{(k)},\mbox{\boldmath$Y$}^{(k)})$
proceeding from the budget constraint $\phi _{k}.\ $ In other words each
individual solves the problem

\begin{eqnarray}
&\max_{\left( \mbox{\boldmath$X$}^{(k)},\mbox{\boldmath$Y$}^{(k)}\right)
}U_{k}(\mbox{\boldmath$X$}^{(k)},\mbox{\boldmath$Y$}^{(k)})&  \label{v2} \\
&\text{if}\quad (\mbox{\boldmath$X$}^{(k)}\cdot \mbox{\boldmath$p$}+%
\mbox{\boldmath$Y$}^{(k)}\cdot \mbox{\boldmath$q$})=\phi _{k}&  \nonumber
\end{eqnarray}
A full consumption level of the goods type $\alpha $ is given by expression

\begin{equation}
X_{\alpha }=\sum_{k}X_{\alpha }^{(k)}  \label{v3}
\end{equation}
The problem~(\ref{v2}) is reduced to solution of the system of equations 
\begin{equation}
\frac{\partial U_{k}(\mbox{\boldmath$Z$}^{(k)})}{\partial Z_{\gamma }^{(k)}}%
=\Lambda _{k}c_{\gamma },  \label{v4}
\end{equation}
where $\Lambda _{k}>0$ are the Lagrange multipliers, the index $\gamma $ runs
through all the set $\{\gamma \}$. The solution of the given system of
equations determines the demand function of the $k$ consumer 
\begin{equation}
\mbox{\boldmath$Z$}^{(k)}=\mbox{\boldmath$S$}^{(k)}(\mbox{\boldmath$c$}%
\Lambda _{k})=(\{S_{\alpha }^{(k)}(\mbox{\boldmath$c$}\Lambda
_{k})\},\{S_{\beta }^{(k)}(\mbox{\boldmath$c$}\Lambda _{k})\}),  \label{v5}
\end{equation}
which depends on both the prices $\mbox{\boldmath$c$}$, and the value $%
\Lambda _{k}$, the latter itself becoming the function of the prices $%
\mbox{\boldmath$c$}$ and of the budget constraints $\phi _{k}$, and satisfies the
equation
\begin{equation}
\mbox{\boldmath$S$}^{(k)}[\mbox{\boldmath$c$}\Lambda _{k}(\mbox{\boldmath$c$}%
,\phi _{k})]\cdot \mbox{\boldmath$c$}=\phi _{k}.  \label{v6}
\end{equation}

The utility function $U_{k}(\mbox{\boldmath$Z$}^{(k)})$ is defined with the
accuracy of monotone transformation $U_{k}\rightarrow {\cal G}(U_{k})$, that
conjugates also with transformation $\Lambda _{k}\rightarrow \Lambda _{k}/%
{\cal G}^{\prime }(U_{k})$. The final type of the function $S^{k}(c,\phi
_{k})$ remains constant. The dependence (\ref{v5}) as the argument contains the
product $\mbox{\boldmath$c$}\Lambda _{k}(\mbox{\boldmath$c$},\phi _{k})$,
where the second multiplier (in his turn, possibly, multiplied on the
function of $1/{\cal G}^{\prime }(U_{k})$ type) actually aggregates in itself
an information about the state of the market of commodities ${\gamma }$ as a
whole and of budget constrains of the consumer $k$. Evident dependence in
this type of function $S_{\gamma }^{(k)}$ on the price $c_{\gamma
^{^{\prime }}}$ with $\gamma \neq \gamma ^{^{\prime }}$ reflects mutual
influence of the goods of sorts $\gamma $ and $\gamma ^{^{\prime }}$ for the
process of their choice. In a considered model a set of goods ${\alpha }$ in
its own right comply with some requirement of the person, while other goods
\{$\beta $\}correspond to other independent requirements. Therefor, it is
natural to expect, that a choice of the set \ $X^{k}$ will be defined only
by prices $\{p\}$ for these goods and by common budget constraint,
aggregated in some uniform multiplicand depending already from all prices $%
\mbox{\boldmath$c$}$ and constrains $\phi _{k}$. Let's note, that the given
assumption is similar to the aggregated description of the utility function
of various commodity groups \cite{Sh87r,VSh91r,Sh93r}, however it has the
difference (see. a chapter~ \ref{va}). Such behavior of the consumers will be
realized, if using some transformation ${\cal {G}}$ the utility function $%
U_{k}$ can be reduced to such type, that for any $\alpha $ and $\beta $

\begin{equation}
\frac{\partial ^{2}U_{k}(\mbox{\boldmath$X$}^{(k)},\mbox{\boldmath$Y$}^{(k)})%
}{\partial X_{\alpha }^{(k)}\partial Y_{\beta }^{(k)}}=0.  \label{v7}
\end{equation}
Taking into account a correlation~(\ref{v7}) we shall accept the following
assumption.

\begin{suggestion}
We suppose, that for the goods $\{\alpha \}$ of the ideal mesoscopic market
for any consumer $k$ there is a utility function $U_{k}(\mbox{\boldmath$Z$}%
^{(k)})$, representable as, 
\begin{equation}
U_{k}(\mbox{\boldmath$X$}^{(k)},\mbox{\boldmath$Y$}^{(k)})=u_{k}(%
\mbox{\boldmath$X$}^{(k)})+v_{k}(\mbox{\boldmath$Y$}^{(k)}).  \label{v8}
\end{equation}
\end{suggestion}
Then, by virtue of~(\ref{v4}) - (\ref{v6}), demand function for the
goods $\mbox{\boldmath$Z$}^{(k)}=(\mbox{\boldmath$X$}^{(k)},%
\mbox{\boldmath$Y$}^{(k)})$ can be rewritten as
\begin{equation}
\mbox{\boldmath$S$}^{(k)}(\mbox{\boldmath$c$},\phi _{k})=(\mbox{\boldmath$S$}%
_{x}^{(k)}(\mbox{\boldmath$p$}\Lambda _{k}),\mbox{\boldmath$S$}_{y}^{(k)}(%
\mbox{\boldmath$q$}\Lambda _{k})),  \label{v9}
\end{equation}
where functions $\mbox{\boldmath$S$}_{x}^{(k)}(\mbox{\boldmath$p$}\Lambda
_{k})=\{S_{\alpha }^{(k)}(\mbox{\boldmath$p$}\Lambda _{k})\}$, $%
\mbox{\boldmath$S$}_{y}^{(k)}(\mbox{\boldmath$q$}\Lambda _{k})=\{S_{\beta
}^{(k)}(\mbox{\boldmath$q$}\Lambda _{k})\}$, satisfy the conditions 
\begin{mathletters}
\begin{eqnarray}
\left. \frac{\partial u_{k}(\mbox{\boldmath$X$}^{(k)})}{\partial X_{\alpha
}^{(k)}}\right| _{\mbox{\boldmath$X$}^{(k)}=\mbox{\boldmath$S$}_{x}^{(k)}}
&=&\Lambda _{k}p_{\alpha },  \label{v10a} \\
\left. \frac{\partial v_{k}(\mbox{\boldmath$Y$}^{(k)})}{\partial Y_{\beta
}^{(k)}}\right| _{\mbox{\boldmath$Y$}^{(k)}=\mbox{\boldmath$S$}_{y}^{(k)}}
&=&\Lambda _{k}q_{\beta },  \label{v10b}
\end{eqnarray}

And the function $\Lambda _{k}(\mbox{\boldmath$p$},\mbox{\boldmath$q$},\phi
_{k})$ is found from the equation 
\end{mathletters}
\begin{equation}
\mbox{\boldmath$S$}_{x}^{(k)}(\mbox{\boldmath$p$}\Lambda _{k})\cdot %
\mbox{\boldmath$p$}+\mbox{\boldmath$S$}_{y}^{(k)}(\mbox{\boldmath$q$}\Lambda
_{k})\cdot \mbox{\boldmath$q$}=\phi _{k}.  \label{v11}
\end{equation}

The value of the $\Lambda _{k}$ parameter is controlled by a state of whole
economic system. Therefore local changes of the prices $\mbox{\boldmath$p$}$
on mesoscopic market of the goods $\{\alpha \}$ should vary the value $%
\Lambda _{k}$. Let's demonstrate it on the example of small fluctuations of
the prices $\delta \mbox{\boldmath$c$}=(\delta \mbox{\boldmath$p$},\delta %
\mbox{\boldmath$q$}).\ $ Varying expression~(\ref{v11}) under condition $%
\phi _{k}=$ {\it const}, we get

\begin{equation}
\sum_{\alpha }(1-\epsilon _{\alpha }^{(k)})S_{\alpha }^{(k)}\delta p_{\alpha
}+\sum_{\beta }(1-\epsilon _{\beta }^{(k)})S_{\beta }^{(k)}\delta q_{\beta
}=(\sum_{\alpha }\epsilon _{\alpha }^{(k)}S_{\alpha }^{(k)}p_{\alpha
}+\sum_{\beta }\epsilon _{\beta }^{(k)}S_{\beta }^{(k)}p_{\beta })\frac{%
\delta \Lambda _{k}}{\Lambda _{k}},  \label{v12}
\end{equation}
where 
\begin{equation}
\epsilon _{\gamma }=-\left. \frac{\partial S_{\gamma }^{(k)}}{\partial
c_{\gamma }}\right| _{\Lambda _{k}=const}  \label{v13}
\end{equation}
From this immediately follows, that, when the number $N$ of independent
mesomarkets is rather great and the fluctuations of the prices is not
significant, so, they do not change \ the state
of the consumer, the mean value ${\left\langle \delta \Lambda _{k}/\Lambda
_{k}\right\rangle \rightarrow 0,\;}${at } $N\rightarrow \infty $ for $1/\sqrt{%
N}$. This leads us to\bigskip

\begin{suggestion}
At the local changes on the market of the goods $\{\alpha \}$ the value $%
\Lambda _{k}$ can be considered as some constant macroscopic variable,
describing common state of the consumer $k$\bigskip 
\end{suggestion}
Taking into account relation~(\ref{v10a}), from this assertion we also
receive, that\bigskip

\begin{suggestion}
At the local changes on the market of the goods $\{\alpha \}$ the strategy
of the consumer $k$ can be described as finding of a maxima of the
following function
\[
\max_{\mbox{\boldmath$X$}^{(k)}}\left[ \frac{1}{\Lambda _{k}}u^{(k)}(%
\mbox{\boldmath$X$}^{(k)})-\mbox{\boldmath$p$}\cdot \mbox{\boldmath$X$}^{(k)}%
\right].
\]
\end{suggestion}

\bigskip In this case function
\[
\frac{1}{\Lambda _{k}}u^{(k)}(\mbox{\boldmath$X$}^{(k)})
\]
can be considered as a utility function of the goods required$%
\mbox{\boldmath$X$}^{(k)}$, depending also from some macroscopic parameters.
This utility function not only defines the preference of a choice, but also
measures this preference in monetary units.

Let$^{\prime }$s now consider the result of collective behavior of
individuals $\{{\cal M}_{k}\}$. A full consumption level $X_{\alpha }$ of
the goods of type $\alpha $ by virtue of~(\ref{v3}) is

\begin{equation}
X_{\alpha }=\sum_{k}S_{\alpha }^{(k)}(\mbox{\boldmath$p$}\Lambda _{k}).
\label{v14}
\end{equation}
Let$^{\prime }$s define the function 
\begin{equation}
{\cal U}(\mbox{\boldmath$p$},\{\Lambda _{k}\})\stackrel{\text{def}}{=}%
\sum_{k}\frac{1}{\Lambda _{k}}u_{k}(\mbox{\boldmath$S$}_{x}^{(k)}(%
\mbox{\boldmath$p$}\Lambda _{k})).  \label{v15}
\end{equation}

If in these two expressions ~(\ref{v14}), (\ref{v15}) we consider values $%
\mbox{\boldmath$p$}$ as some formal set of parameters, than expression~(\ref
{v14}), can be inverted, resulted in some dependence, specifying the
vector $\mbox{\boldmath$p$}$ as the function of a vector $\mbox{\boldmath$X$}%
.\ $ The latter allows us to consider expression ~ (\ref{v15}) as the
function ${\cal U}(\mbox{\boldmath$X$}|\{\Lambda _{k}\})$ of arguments $%
\mbox{\boldmath$X$}$. By virtue of (\ref{v10a}),(\ref{v10b}),(\ref{v14}),(\ref
{v15}) we get

\begin{eqnarray*}
\frac{\partial {\cal U}}{\partial X_{\alpha }} &=&\sum \frac{\partial {\cal U%
}}{\partial p_{\alpha ^{\prime }}}\frac{\partial p_{\alpha ^{\prime }}}{%
\partial X_{\alpha }}=\sum_{k,\alpha ^{\prime },\alpha ^{\prime \prime }}%
\frac{\partial {\cal U}}{\Lambda _{k}\partial S_{\alpha ^{\prime \prime }}}%
\frac{\partial S_{\alpha ^{\prime \prime }}}{\partial p_{\alpha ^{\prime }}}%
\frac{\partial p_{\alpha ^{\prime }}}{\partial X_{\alpha }}=\sum_{k,\alpha
^{\prime },\alpha ^{\prime \prime }}p_{\alpha ^{\prime \prime }}\frac{%
\partial S_{\alpha ^{\prime \prime }}}{\partial p_{\alpha ^{\prime }}}\frac{%
\partial p_{\alpha ^{\prime }}}{\partial X_{\alpha }} \\
&=&\sum_{\alpha ^{\prime },\alpha ^{\prime \prime }}p_{\alpha ^{\prime
\prime }}\frac{\partial p_{\alpha ^{\prime }}}{\partial X_{\alpha }}\frac{%
\partial }{\partial p_{\alpha ^{\prime }}}\left( \sum_{k}S_{\alpha ^{\prime
\prime }}^{(k)}\right) =\sum_{\alpha ^{\prime },\alpha ^{\prime \prime
}}p_{\alpha ^{\prime \prime }}\frac{\partial p_{\alpha ^{\prime }}}{\partial
X_{\alpha }}\frac{\partial X_{\alpha ^{\prime \prime }}}{\partial p_{\alpha
^{\prime }}}=\sum_{\alpha ^{\prime \prime }}p_{\alpha ^{\prime \prime
}}\delta _{\alpha ,\alpha ^{\prime \prime }}=p_{\alpha }
\end{eqnarray*}

Thus, collective behavior of the consumers on the market of the goods \{$%
\alpha $\} can be presented as optimization of the global utility
function of consumption ${\cal U}(\mbox{\boldmath$p$},\{\Lambda _{k}\})$,
which depends also on some macroscopic parameters $\{\Lambda _{k}\}$and is
measured in monetary units. In other words:\bigskip

\begin{suggestion}
\bigskip\ At the local changes on the market of the goods $\{\alpha \}$ the
strategy of cooperative behavior of the consumers can be described as
finding a maximum of the following function: 
\begin{equation}
\max_{\mbox{\boldmath$X$}}\left[ {\cal U}(\mbox{\boldmath$X$}|\{\Lambda \})-%
\mbox{\boldmath$p$}\cdot \mbox{\boldmath$X$}\right] .  \label{v20}
\end{equation}
The common function of demand $\mbox{\boldmath$S$}(\mbox{\boldmath$p$}%
|\{\Lambda \})$ of the goods $\{\alpha \}$ satisfies the equation 
\begin{equation}
\left. \frac{\partial {\cal U}(\mbox{\boldmath$X$}|\{\Lambda \})}{\partial
X_{\alpha }}\right| _{\mbox{\boldmath$X$}=\mbox{\boldmath$S$}}=p_{\alpha }.
\label{vs}
\end{equation}
\end{suggestion}

Measurability of the global utility function in monetary units allows to
compare it with costs of the goods production on the given market and to
describe its operation as a whole in terms of the variational problem. Next
section is devoted to consideration of this problem

\subsection{ Functional of the market efficiency \label{va}}

Let's  use the actually known description of the market
system in terms of global efficiency criterion. Let$^{\prime }$s set a
functional of efficiency ${\cal D}$ of the market goods $\{\alpha \}$ to be:

\begin{equation}
{\cal D}={\cal U}(\{X_{\alpha }\}|\{\Lambda \})-\sum_{i}n_{i}t_{i}(x_{i}),
\label{20}
\end{equation}
Here indexes $\alpha $ and $i$ run over all units of sets ${\cal M}$
and ${\cal N}$, respectively. The arguments of this functional are $%
\{n_{i}\},\{x_{i}\}$,${X_{\alpha }}$ and the structure of the network of
supply ${\cal N}$, which gives conservation laws of material streams, and, hence,
the interrelation \ between values$\{n_{i}\},\{x_{i}\}$, and ${X_{\alpha }}$ values.
The first term in the right part of expression (\ref{20}) is a global
utility of the goods consumption with the level ${X_{\alpha }}$, and the
second term - the production costs .

\begin{suggestion}
The laws of functioning of the ideal mesomarket can be rewritten as the
equations for the extremals of ~(\ref{20}) functional, considering the
network ${\cal N}$, as given.
\end{suggestion}

Indeed, variables $\{n_{i}\},\{x_{i}\}$,${X_{\alpha }}$ are not \
independent, and are connected by relations (\ref{2}) and (\ref{3}). Let's
take advantage of the Lagrange method, which in this case will correspond
to\ the prices $p_{{}_{{\cal B}}}$on the micromarket at node ${\cal B}$

\begin{equation}
\bigskip {\cal D^{P}}={\cal D}+\sum_{\alpha }p_{\alpha }(n_{i_{\alpha
}}x_{i_{\alpha }}-X_{\alpha })\mbox{}+\sum_{{\cal B}}p_{{}_{{\cal B}%
}}(\left. n_{j}x_{j}\right| _{j=i_{in}^{{\cal B}}}-\sum_{j\in \{i_{out}^{%
{\cal B}}\}}n_{j}x_{j}).  \label{21}
\end{equation}
Here index ${\cal B}$ runs over whole nodes (micromarkets) of the
network ${\cal N}$ except nodes, connecting immediately firms and consumers,
and $\{p_{\alpha }\}$ and $\{p_{{}_{{\cal B}}}\}$ are~Lagrange multipliers
(prices) attributed by the nodes network ${\cal N}$. For functional~(\ref{21}%
) arguments $\{n_{i}\}$, $\{x_{i}\}$, $\{X_{\alpha }\}$ and $\{p_{\alpha }\}$
and $\{p_{{}_{{\cal B}}}\}$ are supposed to be independent. As it is well-known,
in this case the extremals of functionals ~(\ref{20}) and (\ref{21})
coincide with each other. Relation~(\ref{21}) can be also rewritten as: 
\begin{equation}
{\cal D^{P}}={\cal U}(\{X_{\alpha }\}|\{\Lambda \})-\sum_{\alpha }p_{\alpha
}X_{\alpha }\mbox{}+\sum_{i}n_{i}\pi _{i}(x_{i}\ ,\ p_{i}^{(s)}-p_{i}^{(b)}),
\label{22}
\end{equation}
where the function $\pi _{i}(\ldots \,,\ \ldots )$ is determined by the
formula~(\ref{4}). Then differentiating the expression ~(\ref{21}) with
respect to variables $\{p_{\alpha }\}$ and $\{p_{{\cal B}}\}$, and
expression~(\ref{22}) with respect to variables $\{n_{i}\}$, $\{x_{i}\}$, $%
\{X_{\alpha }\}$ we get the equations of the model, formulated in the
part~II, where the Lagrange multipliers act as prices, and the demand
function $\mbox{\boldmath$X$}=\mbox{\boldmath$S$}(\mbox{\boldmath$p$}%
|\{\Lambda \})$ satisfies the criterion ~(\ref{vs}).

Let$^{\prime }$s now consider the condition of extremality of a functional~(%
\ref{21})(or(\ref{22}))concerning all the parameters of the order, including
the parameters describing the production and trade system. As follows from
the result of part II, the expression of the functional ${\cal D_{\ast }^{P}}
$, received after maximization over all arguments, except a structure of the
network ~ ${\cal N}$, is determined next formula 
\begin{equation}
{\cal D_{\ast }^{P}}={\cal U}(\{X_{\alpha }\}|\{\Lambda \})-\sum_{\alpha
}f_{\alpha }X_{\alpha }.  \label{23}
\end{equation}
Values $\{f_{\alpha }\}$, and therefore values $\{X_{\alpha }\}$, are given by
production process and geometry of the network ${\cal N}$. If their changes
are such that variations $\{\delta f_{\alpha }\}$, and $\{\delta X_{\alpha
}\}$ are small ,then, by virtue of (\ref{vs}) and (\ref{23}) the variation
of functional ${\cal D_{\ast }^{P}}$ is: 
\begin{equation}
{\delta {\cal D_{\ast }^{P}}}\approx -\sum_{\alpha }X_{\alpha }\delta
f_{\alpha }.  \label{24}
\end{equation}
From (\ref{24}) it follows the comparison criterion of the function
efficiency of ideal markets, identical for their nomenclature of the goods
of consumption $\{\alpha \}$, which differ by organization of the goods
production. Namely:

\begin{suggestion}
Increasing the value of functional efficiency of ideal mesomarket, owing to
the changes of its production and trade structure corresponds to decreasing,
in the average, retail prices for the goods of consumption.
\end{suggestion}

The given result allows, at least on an intuitive level to hypothesize that
for ideal mesomarket the functional of efficiency will have the extremum
including about market structures. Really, it seems to be quite reasonable,
that due to self-organization process only those structures will survive,
which offer the goods with lower prices to the consumers.

The second result, following from the formula(\ref{24}), is, that for the
ideal market exists a plenty of realization of production and trade
structure, which are identical from the viewpoint of extremality of
functional efficiency. In other words:\bigskip

\begin{suggestion}
The ideal market is degenerates concerning the network of supply~${\cal N}$%
\bigskip .
\end{suggestion}

Really, according to expression~(\ref{13}) the amount of parameters,
specifying the value of $\{f_{\alpha }\}$, far exceeds the number
of these values in consequence of hierarchical organization of the network $%
{\cal N}$. That is why the set of values $\{f_{\alpha }\}$ can be obtained by
a large amount of methods. In particular, two realizations of the network $%
{\cal N}$ will result in the single value ${\cal D_{\ast }^{P}}$, if they
contain a single amount of hierarchy levels. Firms, belonging to one level
are characterized by the same function of the costs $t_{i}(x_{i}),$ and
networks are differed only by organization of intermediate bifurcations (Fig.%
\ref{F.2}). In other words, market structures with identical technology,
which have various organization of the micromarket network, are equivalent
on the ideal market.

\section{Remarks\label{sec:rem}}

In the present paper we tried to demonstrate, that the problem of
self-regulation of market systems can be solved through the appearance of
production and trade hierarchical structures. The classical economic theory
and Walras$^{\prime }$s auctioneer suggestion assumes the availability of
the whole information about an economic state of the market. However, it is
physically impossible due to the large number of the participants of the
market system. This limitation can be removed, if the market participants in
their contacts restrict to a small amount of other participants, so that to
form a connected micromarkets network. In doing so, however, occurs another
problem.

Each market participant, contacting with the limited number of others,
receives a small share of information, on which he orients himself planning
and selecting the strategy of his activity. At first glance such limited
information (usually as the prices for some production of the last range)
immediately cannot say him, what and how much to produce. Really, at
availability of the developed system of the micromarkets the result of his
work, usually is an intermediate product (semiproducts). And the market
participant is involved in a long production chain, connecting the
production of raw materials and the final goods offered to the consumers. And
the prices are established during the process of mutual convention of a
small amount of people.

The formulated model of the ideal market with hierarchical structures of the
micromarkets demonstrated the existence of self-regulation process in such systems.
It allows the market participants to adequately react
on changes of the consumers demand, basing only on a value of the
prices on the two appropriate micromarkets, connected together by the
participants. The conservation laws of material flows and streams of money on
the micromarkets are the basis for this self-regulation . Owing to the fact
that \ the stream of money ''flows'' in back direction (from the consumers
to the firms, obtaining raw materials) in contrast to the stream of
materials, the smaller-sized stream of money are going into all bigger-sized
streams. The integration in money streams provide that the
production prices aggregate information about the state and the demand of
the consumers on more and more large scales. The latter provides \ the
self-processing of information, latent in relation of the prices on
semiproducts of different hierarchical levels.

In summary we shall note the following.

First, in an accepted model it was considered, that production and trade
network ${\cal N}$ has a form of a tree. For the market with a perfect
competition such assumption is quite justified. Really, as follows from (\ref
{5}), (\ref{6}), the conclusion about the independence between the $\Delta p_{i}$
difference in prices and demand is not connected with
network ${\cal N}$ geometry, and is the sequence of the perfect competition.
Let assume now, that to some micromarket (node of the network $%
{\cal N}$) has two incoming branches (technological ways), connecting it with
firms, obtaining raw material (root of the network ${\cal N}$ ). Then the
prices for production, proposed on the given micromarket for these two
routes of production, should be different (except accident agreement).
Therefore one of these routes of production has to disappear. And the network $%
{\cal N}$ has to become of a tree form (graph without cycles).

Second, for the ideal market the prices the amount of firms participating in
production and trade, and also the streams of goods appear to be defined
values. The market deviation from ideality, for example, when the
competition is not perfect, and the firms producing the same production
are not identical, will produce the change in these values. However, if
these deviations are not too large, it is possible to expect, that also the
changes in the prices and in the commodity production level also will be
insignificant. The situation varies significantly, concerning market
structures. Owing to a degeneracy of market norm over production and trade
structures, small deviations from perfectness can result in, that during
self-organizing will ''survive'' the structures of various form. Upon it
small variations of parameters unideality can stipulate significant
changes in the market structures. It is likely that, this effect was observed
in paper \cite{GSh95r}. It should be noted, that a plenty of realizations of
production and trade structure, which is identical from the viewpoint of
ideal market, points to the possibility of existence of any macroscopic
parameters of order, characterizing an architecture of these structures in
aggregated form.

In summary to the given part of the paper we note that the assumption about
possibilities of reduction of the common utility function ~(\ref{v7}) is
similar to aggregated description of utility function for the goods,
belonging to different groups \cite{Sh87r,VSh91r,Sh93r}. The main difference
is, that in the proposed approach, budget constrains are not
imposed individually on the goods of each mesomarket and are considered as
common constraints for {\it all} sorts of the goods. As a result, in the
given approach individual indexes of the prices are absent. The description
of the local behavior of the consumers on mesomarket (condition ~(\ref{v20}%
)) in its form is different from the global optimality condition~ (\ref{v2})
. Besides the utility function in the form (\ref{v7}) is uniquely defined,
and that doesn$^{\prime }$t allow beforehand the homogeneity of the first
degree.

We thank to Prof. I. G. Pospelov and Dr. S. Guriev for the useful remarks and
recommendations. Paper also was supported in parts by U1I000 and U1I200 grants
from International Science Fundation, Funds of Fundamental Researches of Russia
Federation and Ukraine.




\setcounter{figure}{0}
\onecolumn
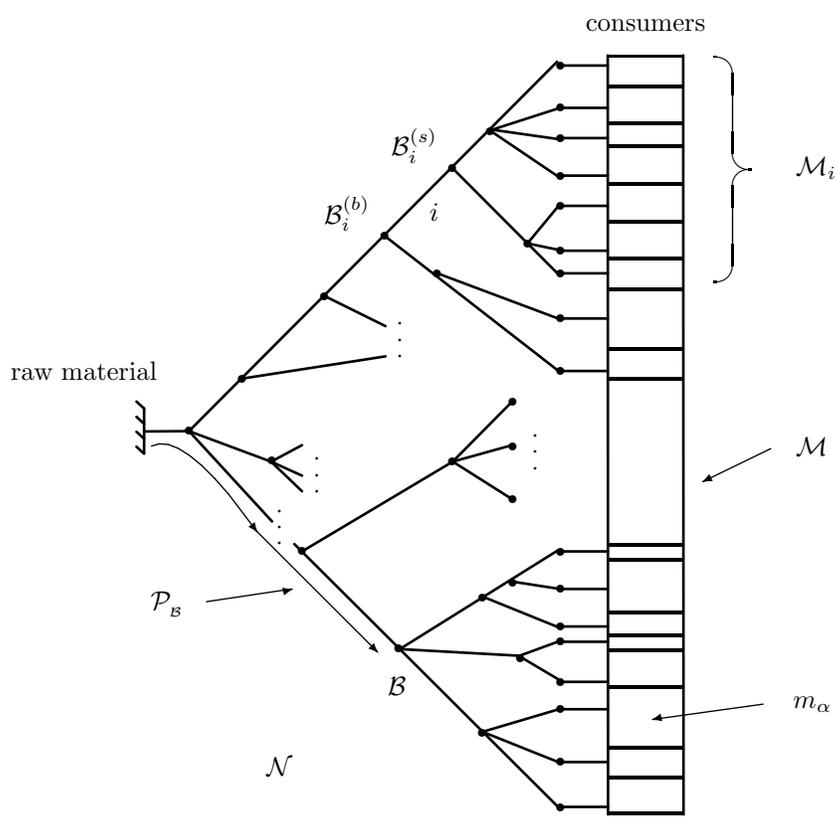
\begin{figure}
\special{em:linewidth 1pt}
\unitlength 1.00mm
\linethickness{1pt}
\begin{picture}(119.67,132.00)(0,-15)
\put(56.00,70.01){\circle*{0.00}}
\put(56.00,72.16){\circle*{0.00}}
\put(56.00,74.31){\circle*{0.00}}
\put(51.00,63.10){\circle*{0.00}}
\put(51.00,65.25){\circle*{0.00}}
\put(51.00,67.40){\circle*{0.00}}
\put(104.67,27.12){\line(-1,0){10.00}}
\put(104.67,127.90){\line(-1,0){10.00}}
\put(39.00,78.13){\circle*{0.90}}
\put(46.00,84.89){\circle*{0.90}}
\put(54.00,62.07){\circle*{0.90}}
\put(73.98,73.99){\circle*{0.90}}
\put(66.86,49.06){\circle*{0.90}}
\put(82.00,57.82){\circle*{0.90}}
\put(77.95,37.92){\circle*{0.90}}
\put(54.01,61.95){\line(0,0){0.00}}
\emline{39.00}{78.00}{1}{33.01}{77.97}{2}
\emline{53.00}{63.00}{3}{87.92}{28.15}{4}
\emline{39.00}{78.00}{5}{87.92}{127.15}{6}
\emline{82.00}{58.00}{7}{88.00}{57.06}{8}
\put(78.00,55.82){\circle*{0.90}}
\emline{67.00}{49.00}{9}{88.00}{62.13}{10}
\emline{67.00}{49.00}{11}{82.94}{48.13}{12}
\emline{82.94}{48.13}{13}{88.00}{49.99}{14}
\emline{83.00}{48.00}{15}{88.00}{45.02}{16}
\emline{78.00}{38.00}{17}{88.00}{40.97}{18}
\put(83.00,47.82){\circle*{0.90}}
\emline{54.00}{62.00}{19}{74.02}{73.95}{20}
\emline{82.00}{82.00}{21}{74.02}{73.95}{22}
\emline{74.02}{73.95}{23}{81.97}{76.13}{24}
\emline{74.00}{74.00}{25}{81.97}{68.97}{26}
\put(81.98,81.99){\circle*{0.90}}
\put(81.98,75.99){\circle*{0.90}}
\put(81.98,68.99){\circle*{0.90}}
\put(78.98,117.99){\circle*{0.90}}
\put(73.98,112.99){\circle*{0.90}}
\put(64.98,103.99){\circle*{0.90}}
\put(56.98,95.99){\circle*{0.90}}
\emline{46.00}{85.00}{27}{65.10}{87.95}{28}
\emline{57.00}{96.00}{29}{65.10}{91.99}{30}
\emline{65.00}{104.00}{31}{88.00}{86.08}{32}
\put(83.98,102.99){\circle*{0.90}}
\emline{84.00}{103.00}{33}{88.00}{107.86}{34}
\emline{79.00}{118.00}{35}{88.00}{111.91}{36}
\emline{79.08}{118.13}{37}{88.00}{116.88}{38}
\emline{39.00}{78.00}{39}{49.91}{73.95}{40}
\emline{49.91}{73.95}{41}{54.00}{76.13}{42}
\emline{54.00}{70.00}{43}{49.91}{73.95}{44}
\emline{49.91}{73.95}{45}{54.00}{72.08}{46}
\put(67.00,88.01){\circle*{0.00}}
\put(67.00,90.16){\circle*{0.00}}
\put(67.00,92.31){\circle*{0.00}}
\emline{78.00}{56.00}{47}{88.05}{51.93}{48}
\emline{84.00}{103.00}{49}{88.05}{102.13}{50}
\emline{74.00}{113.00}{51}{88.05}{98.93}{52}
\put(72.00,116.00){\makebox(0,0)[rc]{${\cal B}_i^{(s)}$}}
\put(63.00,107.00){\makebox(0,0)[rc]{${\cal B}_i^{(b)}$}}
\put(85.00,73.10){\circle*{0.00}}
\put(85.00,75.25){\circle*{0.00}}
\put(85.00,77.40){\circle*{0.00}}
\put(51.00,33.33){\makebox(0,0)[cc]{$\cal N$}}
\put(50.00,74.13){\circle*{0.90}}
\emline{94.67}{128.00}{53}{94.72}{26.98}{54}
\emline{104.67}{128.00}{55}{104.72}{26.98}{56}
\put(104.67,123.90){\line(-1,0){10.00}}
\put(104.67,115.90){\line(-1,0){10.00}}
\put(104.67,110.90){\line(-1,0){10.00}}
\put(104.67,105.90){\line(-1,0){10.00}}
\put(104.67,100.90){\line(-1,0){10.00}}
\put(104.67,96.90){\line(-1,0){10.00}}
\put(104.67,88.90){\line(-1,0){10.00}}
\put(104.67,84.90){\line(-1,0){10.00}}
\put(104.67,62.90){\line(-1,0){10.00}}
\put(104.67,60.90){\line(-1,0){10.00}}
\put(104.67,53.90){\line(-1,0){10.00}}
\put(104.67,50.90){\line(-1,0){10.00}}
\put(104.67,48.90){\line(-1,0){10.00}}
\put(104.67,43.90){\line(-1,0){10.00}}
\put(104.67,35.90){\line(-1,0){10.00}}
\put(104.67,31.90){\line(-1,0){10.00}}
\put(71.00,106.00){$i$}
\put(119.67,76.00){\makebox(0,0)[lc]{${\cal M}$}}
\put(119.67,113.00){\makebox(0,0)[lc]{${\cal M}_i$}}
\put(99.67,132.00){\makebox(0,0)[cc]{consumers}}
\emline{33.00}{81.00}{57}{33.02}{75.07}{58}
\emline{33.00}{75.00}{59}{32.02}{75.93}{60}
\emline{33.00}{77.00}{61}{32.02}{77.93}{62}
\emline{33.00}{79.00}{63}{32.02}{79.93}{64}
\emline{33.00}{81.00}{65}{32.02}{81.93}{66}
\put(25.00,86.00){\makebox(0,0)[cc]{raw material}}
\emline{79.00}{118.00}{67}{88.05}{120.95}{68}
\put(104.67,118.90){\line(-1,0){10.00}}
\put(71.98,98.99){\circle*{0.90}}
\emline{72.00}{99.00}{69}{88.05}{92.98}{70}
\emline{78.00}{38.00}{71}{87.96}{34.04}{72}
\emline{39.00}{78.00}{73}{50.03}{65.98}{74}
\put(88.33,126.67){\circle*{0.94}}
\emline{88.33}{126.67}{99}{94.67}{126.67}{100}
\put(88.33,121.00){\circle*{0.94}}
\emline{88.33}{121.00}{101}{94.67}{121.00}{102}
\put(88.33,117.00){\circle*{0.94}}
\emline{88.33}{117.00}{103}{94.67}{117.00}{104}
\put(88.33,112.00){\circle*{0.94}}
\emline{88.33}{112.00}{105}{94.67}{112.00}{106}
\put(88.33,108.00){\circle*{0.94}}
\emline{88.33}{108.00}{107}{94.67}{108.00}{108}
\put(88.33,102.00){\circle*{0.94}}
\emline{88.33}{102.00}{109}{94.67}{102.00}{110}
\put(88.33,99.00){\circle*{0.94}}
\emline{88.33}{99.00}{111}{94.67}{99.00}{112}
\put(88.33,93.00){\circle*{0.94}}
\emline{88.33}{93.00}{113}{94.67}{93.00}{114}
\put(88.33,86.00){\circle*{0.94}}
\emline{88.33}{86.00}{115}{94.67}{86.00}{116}
\put(88.33,62.00){\circle*{0.94}}
\emline{88.33}{62.00}{117}{94.67}{62.00}{118}
\put(88.33,57.00){\circle*{0.94}}
\emline{88.33}{57.00}{119}{94.67}{57.00}{120}
\put(88.33,52.00){\circle*{0.94}}
\emline{88.33}{52.00}{121}{94.67}{52.00}{122}
\put(88.33,50.00){\circle*{0.94}}
\emline{88.33}{50.00}{123}{94.67}{50.00}{124}
\put(88.33,44.67){\circle*{0.94}}
\emline{88.33}{44.67}{125}{94.67}{44.67}{126}
\put(88.33,41.00){\circle*{0.94}}
\emline{88.33}{41.00}{127}{94.67}{41.00}{128}
\put(88.33,34.00){\circle*{0.94}}
\emline{88.33}{34.00}{129}{94.67}{34.00}{130}
\put(88.33,28.00){\circle*{0.94}}
\emline{88.33}{28.00}{131}{94.67}{28.00}{132}
\special{em:linewidth 0.5pt}
\linethickness{0.5pt}
%
\put(107.34,71.33){\vector(-2,-1){0.2}}
\emline{116.34}{75.67}{75}{107.34}{71.33}{76}
\put(119.33,41.67){\makebox(0,0)[lc]{$m_\alpha$}}
\put(100.67,39.67){\vector(-4,-1){0.2}}
\emline{115.33}{41.67}{77}{100.67}{39.67}{78}
\put(66.67,45.33){\makebox(0,0)[ct]{${\cal B}$}}
\put(48.00,64.67){\vector(3,-4){0.2}}
\emline{34.00}{76.00}{79}{35.05}{76.32}{80}
\emline{35.05}{76.32}{81}{36.24}{76.23}{82}
\emline{36.24}{76.23}{83}{37.57}{75.73}{84}
\emline{37.57}{75.73}{85}{39.04}{74.81}{86}
\emline{39.04}{74.81}{87}{40.64}{73.49}{88}
\emline{40.64}{73.49}{89}{42.39}{71.74}{90}
\emline{42.39}{71.74}{91}{44.27}{69.59}{92}
\emline{44.27}{69.59}{93}{48.00}{64.67}{94}
\put(64.00,48.67){\vector(1,-1){0.2}}
\emline{48.00}{64.67}{95}{64.00}{48.67}{96}
\put(38.33,55.00){\makebox(0,0)[rc]{${\cal P}_{{}_{\cal B}}$}}
\put(52.67,57.00){\vector(4,1){0.2}}
\emline{41.33}{55.33}{97}{52.67}{57.00}{98}
\put(108.74,122.74){\oval(5.02,10.04)[rt]}
\put(113.76,117.77){\oval(5.02,9.94)[lb]}
\emline{111.25}{123.79}{133}{111.25}{116.78}{134}
\put(108.74,102.86){\oval(5.02,10.04)[rb]}
\put(113.76,107.83){\oval(5.02,9.94)[lt]}
\emline{111.25}{101.82}{135}{111.25}{108.83}{136}
\end{picture}
\caption{Structure of the mezomarket under consideration.}
\label{F.1}
\end{figure}

\begin{figure}
\special{em:linewidth 0.8pt}
\unitlength 0.5mm
\linethickness{0.4pt}
\begin{picture}(104.67,82.00)(-80,0)
\put(45.00,30.00){\circle*{2.98}}
\put(42.33,30.00){\vector(1,0){0.2}}
\emline{5.00}{30.00}{1}{42.33}{30.00}{2}
\put(85.00,30.00){\vector(1,0){0.2}}
\emline{45.00}{30.00}{3}{85.00}{30.00}{4}
\put(85.00,50.00){\vector(2,1){0.2}}
\emline{45.00}{30.00}{5}{85.00}{50.00}{6}
\put(85.00,10.00){\vector(2,-1){0.2}}
\emline{45.00}{30.00}{7}{85.00}{10.00}{8}
\special{em:linewidth 0.4pt}
\emline{89.00}{52.00}{9}{90.89}{51.59}{10}
\emline{90.89}{51.59}{11}{92.43}{50.77}{12}
\emline{92.43}{50.77}{13}{93.62}{49.53}{14}
\emline{93.62}{49.53}{15}{94.47}{47.87}{16}
\emline{94.47}{47.87}{17}{94.97}{45.79}{18}
\emline{94.97}{45.79}{19}{95.00}{41.00}{20}
\emline{101.00}{30.00}{21}{99.11}{30.41}{22}
\emline{99.11}{30.41}{23}{97.57}{31.23}{24}
\emline{97.57}{31.23}{25}{96.38}{32.47}{26}
\emline{96.38}{32.47}{27}{95.53}{34.13}{28}
\emline{95.53}{34.13}{29}{95.03}{36.21}{30}
\emline{95.03}{36.21}{31}{95.00}{41.00}{32}
\emline{89.00}{8.00}{33}{90.89}{8.41}{34}
\emline{90.89}{8.41}{35}{92.43}{9.23}{36}
\emline{92.43}{9.23}{37}{93.62}{10.47}{38}
\emline{93.62}{10.47}{39}{94.47}{12.13}{40}
\emline{94.47}{12.13}{41}{94.97}{14.21}{42}
\emline{94.97}{14.21}{43}{95.00}{19.00}{44}
\emline{101.00}{30.00}{45}{99.11}{29.59}{46}
\emline{99.11}{29.59}{47}{97.57}{28.77}{48}
\emline{97.57}{28.77}{49}{96.38}{27.53}{50}
\emline{96.38}{27.53}{51}{95.53}{25.87}{52}
\emline{95.53}{25.87}{53}{95.03}{23.79}{54}
\emline{95.03}{23.79}{55}{95.00}{19.00}{56}
\put(47.33,35.33){\makebox(0,0)[rb]{${\cal B}$}}
\put(20.00,34.00){\makebox(0,0)[cb]{$i_{in}^{\cal B}$}}
\put(104.67,30.00){\makebox(0,0)[lc]{$\left\{i_{out}^{\cal B}\right\}$}}
\end{picture}
\caption{Structure of an elementary micromarket.}
\label{F.a}
\end{figure}
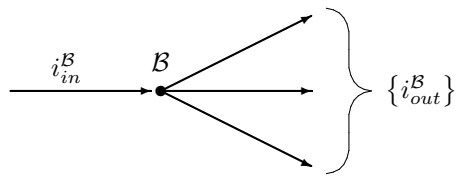

\begin{figure}
\special{em:linewidth 1pt}
\unitlength 0.35mm
\linethickness{1pt}
\begin{picture}(312.67,152.00)(-50,0)
\put(30.00,120.00){\circle*{4.00}}
\put(30.00,120.00){\circle*{4.00}}
\put(210.67,120.00){\circle*{4.00}}
\put(210.67,120.00){\circle*{4.00}}
\put(80.00,140.00){\circle*{4.00}}
\put(80.00,140.00){\circle*{4.00}}
\put(130.00,150.00){\circle*{4.00}}
\put(130.00,130.00){\circle*{4.00}}
\put(130.00,110.00){\circle*{4.00}}
\put(130.00,90.00){\circle*{4.00}}
\put(310.67,150.00){\circle*{4.00}}
\put(310.67,130.00){\circle*{4.00}}
\put(310.67,110.00){\circle*{4.00}}
\put(310.67,90.00){\circle*{4.00}}
\put(79.67,100.00){\circle*{4.00}}
\put(260.34,120.00){\circle*{4.00}}
\emline{210.67}{120.00}{1}{260.34}{120.00}{2}
\emline{260.34}{120.00}{3}{310.67}{150.00}{4}
\emline{310.67}{130.00}{5}{260.67}{120.00}{6}
\emline{260.34}{120.00}{7}{310.67}{90.00}{8}
\emline{310.67}{110.00}{9}{260.67}{120.00}{10}
\emline{30.00}{120.00}{11}{80.00}{140.00}{12}
\emline{80.00}{140.00}{13}{130.00}{150.00}{14}
\emline{130.00}{130.00}{15}{80.00}{140.00}{16}
\emline{30.00}{120.00}{17}{80.00}{100.00}{18}
\emline{80.00}{100.00}{19}{130.00}{90.00}{20}
\emline{130.00}{110.00}{21}{80.00}{100.00}{22}
\special{em:linewidth .5pt}
\linethickness{.5pt}
\emline{170.00}{125.00}{23}{185.00}{125.00}{24}
\emline{192.67}{120.00}{25}{183.00}{130.33}{26}
\emline{170.00}{115.00}{27}{185.00}{115.00}{28}
\emline{192.67}{120.00}{29}{183.00}{109.67}{30}
\emline{170.00}{125.00}{31}{155.00}{125.00}{32}
\emline{147.33}{120.00}{33}{157.00}{130.33}{34}
\emline{170.00}{115.00}{35}{155.00}{115.00}{36}
\emline{147.33}{120.00}{37}{157.00}{109.67}{38}
\end{picture}
\caption{Equivalent fragments of the production-trade network.}
\label{F.2}
\end{figure}
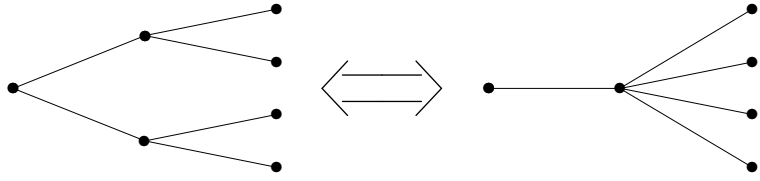


\newpage

\begin{references}
\bibitem{NP79r}  Nikolis G. , Prigogin I.
\newblock {\em Self-organization in nonequilibrum systems. From dissipativ
structures to the order  through fluctuations}
\newblock {(J.Wiley and Sons, 1977).}

\bibitem{H80r}  Haken H.
\newblock {\em  Synergetics. An
introduction. Springer Ser.in Synergetics. v.1}
\newblock {(Springer-verlag,
1988).}

\bibitem{H91r}  Haken H.
\newblock {\em  Information and self-organizing. A macroscopic approach
  to complex systems} \newblock {(Springer-verlag, 1988).}

\bibitem{K95r}  Klimontovich ~Yu. ~L.
\newblock {\em Statistical theory of
open systems} \newblock {(Janus, Moscow, 1995).}

\bibitem{ker}  Kerner B. S. , Osipov V. V.
\newblock {\em Autosolitons. A new approach to
problem of self-organozation and turbulance}
\newblock {(Kluwer, 1994).}


\bibitem{H92r}  Hayek F. A.
\newblock {\em The fatal conceint. The errors of
socialism} \newblock {(The Univ. Chicago press, 1988).}

\bibitem{W91}  Weidlich W.
\newblock {\em Physics and social science -- the
approach of synergetics}
\newblock {// Phys.~Rep.~1991, v.~204, n.~1,
pp.~1--163.}


\bibitem{B77}  Baesemann R. C.
\newblock {\em The formation of small market places in a competitive economic
  process -- the dynamics of agglomeration}
\newblock {//
Econometrica,~1977, v.~45, n.~2, pp.~361--376.}

\bibitem{RW85}  Rubinstein A. , Wolinsky A.
\newblock {\em Equilibrium in a
market with sequential bargaining}
\newblock {// Econometrica~1985, v.~53,
n.~5, pp.~1133--1150.}

\bibitem{W90}  Wolinsky A.
\newblock {\em Information revelatio in a market
with pairwise meetings}
\newblock {// Econometrica~1990, v.~58, n.~1,
pp.~1--24.}

\bibitem{RG94}  van Raalte C. L. , Gilles R. P.
\newblock {\em Endogeneous Formation of Trade Center: An Evolutionary
  Approach}
\newblock {(Department of Economics and CentER, Tilburg
University, 1994).}

\bibitem{GSh95r}  Guriev S. ,Shahova M.
\newblock {\em Model of self-organizing of commercial
      networks in economy  with incomplete infrastructure} \newblock
{In: Mathematical models of the analysis of data and management in
microeconomics }
\newblock{ (Moscow, 1995).}

\bibitem{M89r}  Mchedeshvili G.
\newblock {\em Microcirculation of blood: common features of regulation and
     violations} \newblock { (Science, Leningrad, 1989).}

\bibitem{LG95ES}  Lubashevsky I. A. , Gafiychuk V. V.
\newblock {\em A simple
model of self-regulation in large naturla systems}
\newblock {//
J.~Envir.~Syst.~1995, v.~23, n.~3, pp.~281--289.}

\bibitem{gaf1}  Gafiychuk V. V. , Lubashevsky I. A. \newblock{\em On hierarchical
Structures Arising Spontaneously in Markets with Perfect Competition}
\newblock {//J. Env. Systems. - v. 25, No. 2. - p. 159-166. (1996-1997).}

\bibitem{gaf2}  Lubashevsky I. A. , Gafiychuk V. V. \newblock{\em Cooperative mechanism of
self-regulation in hierarchical living systems} e- print. http://xxx. lanl.
gov/abs/adap-org/9808003 (1998), submitted to SIAM J. Appl. Math.

\bibitem{P88r}  Pospelov I. G. \newblock {\em  Dynamic Model of the Market} %
\newblock {// Economy and Math. Methods, 1988, v.~ 24, no. 3.}

\bibitem{P89r}  Pospelov I. G.
\newblock {\em Optimal strategy of behaviour in the dynamic market
        models}
\newblock {// An Automation and telemechanics, 1989, no.2,
pp. 113 - 123.}

\bibitem{BH88ec}  B. ~R. Binger and E. ~Hoffman.
\newblock {\em Microeconomics
with Calculus} \newblock {(Harper Collins Publishers, 1988).}

\bibitem{SR90}  Scherer F. M. , Ross D.
\newblock {\em Industrial Market
Structure and Economic Performance}
\newblock {(Houghton Mifflin Company,
Boston, 1990).}

\bibitem{Sh87r}  Shananin A. A.
\newblock {\em A Condition of integrability
in the problem of finite goods}
\newblock {// Rep. AS USSR, 1987, v.~ 294,
no.~ 3, pp. ~ 553 - 555.}

\bibitem{VSh91r}  Vratenkov S. D. ,Shananin A. A.
\newblock {\em  Analysis of a structure of consumer demand with
           economic indexes}
\newblock {(Computer Center AS USSR, Moscow,
1991).}

\bibitem{Sh93r}  Shananin A. A.
\newblock {\em A nonparametric method of the analysis of the
consumer's demand structure}
\newblock {// Math. Simulation, 1993, v.~ 5,
no.~ 9, pp.~ 3 - 17.}
\end{references}
\end{document}